\documentclass[sigconf]{acmart}

\setcopyright{none}

\acmDOI{}

\acmISBN{}

\acmConference{}{}{}
\acmYear{}
\copyrightyear{}

\acmArticle{}
\acmPrice{}

\usepackage{amsmath}
\usepackage{booktabs}       
\usepackage{todonotes}
\usepackage{array}
\usepackage{subcaption}
\usepackage{bbm}
\begin{document}

\title{Reacting to Variations in Product Demand: An Application for Conversion Rate (CR) Prediction in Sponsored Search}

\author{Marcelo Tallis}
\affiliation{%
   \institution{Criteo Labs,}
  \streetaddress{325 Lytton Ave. Suite 200}
  \city{Palo Alto} 
  \state{CA}
  \postcode{94301}  
  \country{USA}
}
\email{m.tallis@criteo.com}

\author{Pranjul Yadav}
\affiliation{%
   \institution{Criteo Labs,}
  \streetaddress{325 Lytton Ave. Suite 200}
  \city{Palo Alto} 
  \state{CA}
  \postcode{94301}  
  \country{USA}
}
\email{p.yadav@criteo.com}


\begin{abstract}
In online internet advertising, machine learning models are widely used to compute the likelihood of a user engaging with product related advertisements. However, the performance of traditional machine learning models is often impacted due to variations in user and advertiser behavior. For example, search engine traffic for florists usually tends to peak around Valentine's day, Mother's day, etc. To overcome, this challenge, in this manuscript we propose three models which are able to incorporate the effects arising due to variations in product demand. The proposed models are a combination of product demand features, specialized data sampling methodologies and ensemble techniques. We demonstrate the performance of our proposed models on datasets obtained from a real-world setting. Our results show that the proposed models more accurately predict the outcome of users interactions with product related advertisements while simultaneously being robust to fluctuations in user and advertiser behaviors.
\end{abstract}

%
%



\maketitle

\def\x{\mathbf x}
\def\w{\mathbf w}

\section{Introduction} \label{sec:introduction}
Digital advertising can be performed in multiple forms i.e. contextual advertising, display based advertising and search-based advertising \cite{zeff1999advertising}. In this manuscript we are interested in the promotion of products through a search-based advertising service. Search based advertising has been of great interest considering the numbers of products sold worth million of dollars in a year \cite{king2007internet}. In such advertising, textual/image based advertisements are placed next to their search results and performance is evaluated with a cost-per-click (CPC) billing, which implies that the search engine is paid every time the advertisement is clicked by a user. The search engine typically matches products which are close to the intent of the user (as expressed by the search query) and select products which have the highest bid. 

From an advertisers standpoint, an effective advertisement bidding strategy requires determining the probability that an advertisement click will originate a conversion. Conversion can be defined as either a sale of the product or some corresponding action (i.e. filling of forms or watching a video). Development of effective advertisement bidding strategies require sophisticated machine learning models to compute the likelihood of a user engaging with product related advertisements. Such machine learning models are typically developed using features drawn from a variety of sources: product information consisting of product categories, type, price, age, gender, prior user information including whether a user is new customer vs returning customer, and attributes of the search request including day, time and client device type.

Typically, Generalized Linear Models (GLMs) \cite{mccullagh1984generalized} are widely used to model conversion prediction tasks for multiple reasons: (1) Predictions for billions of search products are made on a daily basis and hence the inference or the likelihood of the user engagement for any product needs to be computed within a fraction of a second (2) GLMs facilitate a quick update of the model parameters as newer data is available. (3) Extreme sparsity of the data i.e. minute fraction of nonzero feature values per data point.

However, traditional GLMs are often limited in their ability to model variations in product demand, originating either due to user buying behavior or advertiser selling behavior (sales). Example of such variations in product demand include : Increase in demand for online florists around Valentine's day, Mother's day, etc., the introduction of a new product into the market (e.g., a new iphone model) or even a competitor that decided to lower some prices. 

The problem becomes relevant when there is a surge in product demands as advertisers miss opportunities because the machine learning models often under-predict user engagement levels with respect to product advertisements. Similarly, when there is a drop in product demand, advertisers overspend because the models over-predict user responses to product advertisements. This usually stems from the fact that traditional machine learning models are slow to incorporate the sudden variations arising due to increase in user buying behavior or advertiser selling behavior. Such biased models in turn lead to significant revenue loss.  

To overcome the challenges associated with variations in product demand, in this manuscript, we propose three novel approaches: Firstly, we extend the baseline model by adding novel features which capture variations in product demand. Secondly, we propose models in conjunction with importance weighting \cite{vasile2016cost}, wherein data from the recent past is weighted more as compared to data from the distant past during model training, and lastly we propose mixture models i.e. models consisting of our original model along with a model trained on more recent data. We then demonstrate the superior performance of our proposed models w.r.t the baseline model in a real world setting.

\section{Related Work}

Several interesting machine learning research have been performed in the domain of contextual advertising, sponsored search advertising \cite{fain2006sponsored} and display advertising \cite{mcmahan2013ad, chapelle2015simple}. Current state-of-the-art conversion rate (CR) prediction methods range from logistic regression \cite{chapelle2015simple}, to log-linear models \cite{agarwal2010estimating} and to a combination of log-linear models with decision trees \cite{he2014practical}. More complex modeling techniques such as deep learning, ensemble methods and factorization machines have also been widely used for such tasks. 

Within the realm of deep learning, Zhang et al. \cite{zhang2014sequential} modeled the dependency on user's sequential behaviors into the click prediction process through Recurrent Neural Networks (RNN). They further concluded that using deep learning techniques led to significant improvement in the click prediction accuracy. On similar lines, Jiang et al. \cite{jiang2016research} proposed deep architecture model that integrates deep belief networks (DBN) with logistical regression to deal with the problem of CTR prediction for contextual advertising. In their work, they used DBN to generate non-linear embeddings from original data (users' information, click logs, product information and pages information) and then they used these embeddings as features into a regression model for CTR prediction problems.

Ensemble models have also been widely used for CTR prediction problems. In particular, He et al. \cite{he2014practical} proposed a model which combines decision trees with logistic regression and observed that the joint model outperforms either of these methods thereby leading to a significant impact to the overall system performance. They concluded that the superior performance was a result of utilizing the right features i.e. those capturing historical information about the user or the advertisement.  Juan et al.  \cite{juan2016field} proposed Field aware Factorization Machines for classifying large sparse data. They hypothesized that feature interactions seems to be crucial for Click-Through-Rate (CTR) predictions and discussed how degree-2 polynomial mappings and factorization machines \cite{rendle2010factorization} handle feature intersections. Richardson et al.\cite{richardson2007predicting} proposed the use of features comprising of information about product advertisements, product description, and advertisers to learn a model that accurately predicts the CTR for new advertisements. 

To explore the effect of keyword queries on CTR prediction, Regelson et al. \cite{regelson2006predicting} hypothesized that keyword terms have a different likelihood of being clicked and hence proposed a novel CTR prediction algorithm to reflect the role of keyword queries. Their algorithm consisted of clusters comprising on keyword terms and observed that clustered historical data leads to accurate CTR estimation. To analyze the effect of other factors towards CR prediction, Chen and Yan \cite{chen2012position} proposed a probabilistic factor model to study the impact of position bias under the hypothesis that higher positions advertisements usually get more clicked and Cheng et al. \cite{cheng2010personalized} proposed user-specific and demographic-based features that reflect the click behavior of individuals and groups and hence proposed a framework for the personalization of click models in sponsored search.

\begin{figure}[h]
	\includegraphics[width=0.5\textwidth,height=0.5\textwidth]{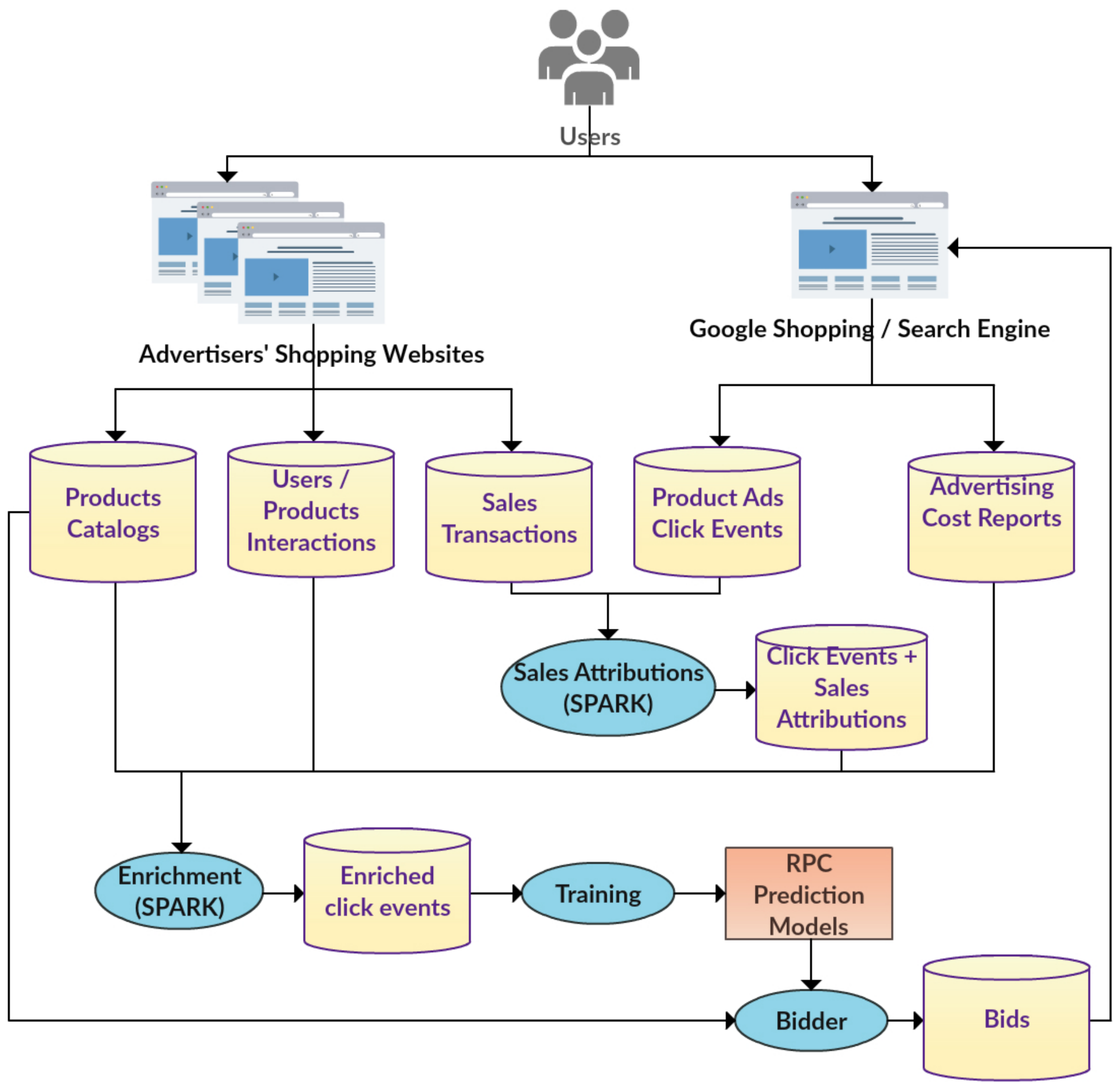}
    \centering
    \caption{Criteo Predictive Search Data Processing Pipeline}
    \label{fig:data_procesing}
\end{figure}

\section{Criteo Predictive Search}
\label{sec:CPS}
Criteo Predictive Search (CPS) is a recent product from Criteo that was launched at the end of 2016 on selected markets. CPS uses machine learning to automate all aspects of Google Shopping Campaign optimization i.e. bidding, retargeting lists and product matching. 

When a Google Shopping user searches for products, Google Shopping selects a list of product advertisements to be embedded in the search results page. Google determines which product advertisements to be displayed to the user by weighing how closely the advertised product matches the user's query intent, the probability that an advertisement will be clicked by the user, and the bid amount by the advertiser. Google will only charge (second-price auction) for a displayed advertisement if that advertisement is clicked by a user.

CPS combines different types of data to train machine learning models for predicting the \emph{return per ad-click (RPC)}.  CPS uses these models to produce daily contextualized bids for every product in an advertiser's product catalog. These bids are conditioned by different contexts which are determined by the advertised products, users classes, device types, and negative-keywords query filters. The computed bids are optimized to maximize an advertiser's \emph{return-on-investment (ROI)} under given cost constraints. 

Figure \ref{fig:data_procesing}, depicts the CPS data processing pipelines. CPS collects data from two sources. From advertisers it collects anonymized data about users interacting with advertiser's product pages. Also from advertisers it collects detailed data about sales transactions, including information about the buying user, the products being sold and the amount paid. 
From Google Shopping an Adwords, CPS collects data about ad-click events, including the advertised product, the clicking user, and contextual information (e.g., device type). Also from AdWords it collects daily advertisers' cost reports, which consists on advertisement costs aggregated by product.

A process called \emph{attribution}, aims to match sales transactions with the corresponding ad-click event if any. The purpose of sales attribution is to determine the total return attributed to an ad-click, which can be zero if an ad did not originated any sales. The average return-per-click (RPC) is the target that our ML models attempt to predict. 

Click events together with their attributed sales information are enriched with product related information from advertisers' catalogs (e.g., product price, product category) and with statistics about users / products interactions collected from advertisers' shopping websites (e.g., the number of pages views in the last week for a particular product). Click events data is also enriched with average cost-per-click information computed from  AdWords cost reports.

The enriched click events data is used to train a series of machine learning models needed to predict return-per-clicks (RPC). These models are later used by the \emph{Bidder} to compute the following's day optimal bid for every product in the advertisers' catalogs.
\section{Background}
In this section, we would discuss the baseline technique used to model the likelihood of the user engagement with a product advertisement along with the metrics used to evaluate the performance of the models (i.e. proposed models along with the baseline model). Next, we would provide an overview of the real world dataset used to evaluate the performance of our proposed models. Lastly, we discuss the longitudinal model evaluation protocol followed in this manuscript.
\subsection{Baseline Model} 
We use L2-regularized logistic regression as our baseline approach to model the likelihood of a user engaging with product related advertisement. Specifically, the likelihood of engagement $E(y_i/x_i)$, can be expressed via the link function of a GLM (logistic regression) as follows :

\begin{eqnarray}
E(y_i/x_i)  = \frac{1}{1 + \exp (- \w\cdot\x_i)}
\end{eqnarray} 

The model coefficients $\w$ can be obtained by minimizing the $L_2$ regularized logistic loss, denoted by \emph{Negative Log-Likelihood (NLL)}, as defined below:-

\begin{eqnarray}
\text{NLL} = \arg\min_\w \sum_{i=1}^N \log(1+\exp(-y_i\w\cdot\x_i)) + \frac{\lambda}{2}\|\w\|^2,
\end{eqnarray} 

where in, 
\begin{itemize}
  \item $\lambda$ denotes the regularization parameter.
  \item $w$ denotes the coefficient vector.
  \item $y_i$ denotes the label (1 indicates conversion and -1 indicates no-conversion).
  \item $x_i$ denotes the covariates comprising of information about the product (type, gender, age, attributes, description, color), user(new vs old), etc.
\end{itemize}

L2-regularized logistic regression was chosen as our baseline approach, as it is quick to update model parameters in the presence of new data instances, billions of inferences can be computed in a reasonable amount of time and the model can be trained on massively sparse datasets comprising millions of explanatory variables. L2-regularized logistic regression is also the method being used currently in production.

\subsection{Metrics}
\subsubsection{ \textbf{Log Likelihood Normalized (LLHN)}}
We want our models to predict the probability that an advertisement click will end up in a sale or a conversion. To evaluate our models we want to measure how far the predicted probabilities are from the actual conversion probabilities. Conventional metrics like accuracy, precision, recall and F1-Score, which are usually used to evaluate classification systems, are not sufficiently precise for our purposes. Instead, we would like to rely on metrics related to the \emph{likelihood} of the test data under the evaluated models. 

Given a dataset and a model the likelihood is the probability to observe this dataset assuming the model is true.

Here our dataset is denoted by pair $(x_i, y_i)$ for $i=1..n$ where the $x_i \in R^d $ are the features vectors and the $y_i \in \{-1,1\}$ are the labels.

We assume that the $y_i$ are independent conditionally on the $x_i$ so we can write

$$ Likelihood = \prod_{i=1}^n P(y_i ~|~ x_i) $$
where $P(y_i ~|~ x_i)$ is the probability predicted by a model.

As the likelihood is positive and that the logarithm is an increasing function, maximizing the likelihood is the same as maximizing the log-likelihood (LLH), and the log-likelihood is a sum which is more practical. 

$$ LLH (model) = \sum_{i=1}^n \log(1+\exp(-y_i\w\cdot\x_i)) $$

\emph{Log-likelihood} (LLH), as a metric, is not normalized and hence it cannot be used to compare performances based on different datasets. To overcome this limitation we introduce \emph{Log-Likelihood-Normalized} which is denoted as \emph{LLHN}. The LLHN of model corresponds to its LLH relative to the LLH of a naive model. 

The naive model is the model that predicts a constant '$c$' which is the probability that $y_i=1$ on the test dataset.  \\
$$ c = \frac{1}{n}\sum_{i=1}^n \mathbbm{1}{(y_i = 1)}$$
 and the naive model (naive) does 
$$P(y_i=1) = c$$

The LLH(naive) of the naive model is defined as:-

$$ LLH (naive) = \sum_{i=1}^n (\frac{(1+y_i)log(c)}{2} + \frac{(1 - y_i)log(1-c)}{2} ) $$

The LLHN is defined as :-
$$ LLHN (model) = \frac{LLH(naive)-LLH(model)}{LLH(naive)} $$

A positive LLHN indicates that our model is better than the naive model. For example, a LLHN value of 0.13 means that our model is 13\% "better" than the naive model. A negative LLHN indicates that our model is worse than the naive model. A zero LLHN indicates that our model is equal to the naive model. Ideally, we would like the LLHN value to be as high as possible.

\subsubsection{ \textbf{LLHN-Uplift}} The LLHN-Uplift of a model can be defined as the following :

$$ LLHN-Uplift (model) = \frac{LLHN(model)-LLHN(baseline)}{LLHN(baseline)} $$
 
 where in,
 
\begin{itemize}
  \item LLHN(model) denotes the log-likelihood of any proposed model w.r.t naive model.
  \item LLHN(baseline) denotes the log-likelihood of the baseline model (L2-regularized logistic regression) w.r.t the naive model.
\end{itemize}

A positive LLHN-Uplift indicates that our model is better than the baseline model. For example, a LLHN-Uplift value of 0.15 means that our model is 15\% "better" than the baseline model. A negative LLHN-Uplift indicates that our proposed model is worse than the baseline model. A zero LLHN-Uplift indicates that our model is equal to the baseline model.

\subsection{Data}
We train our models from CPS traffic logs of click events which combine several sources of information, including attributes of the product being advertised (e.g., product id, price, category, brand, retailer), user shopping behavior information (e.g., engagement level),  attributes of the click event (e.g., event time, device), and advertisement performance information (e.g., number of sales and revenue generated). To train models for predicting conversion probabilities we label each event with the event outcome. That is whether that event originated in a conversion or not.

Currently, these logs include several thousand daily events on average. Our models are trained, several times within a day to predict conversion probabilities for millions of advertised products. To train a model to predict conversion probabilities for each product on a particular day we include event's data spanning a couple of days prior to the day whose conversion probabilities we want to predict. Sample CPS dataset has also been publicly released (http://research.criteo.com/criteo-sponsored-search-conversion-log-dataset/).

\subsection{Longitudinal Model Evaluation}
In this manuscript, we simulate the conditions from the production setting. In the production setting, models are trained everyday to make prediction based on most recent data. Then, for each day in our test dataset we train a corresponding model using only data from a period (last couple of days or weeks) preceding the test day. We then compare the outcome of each test event with the outcome predicted by our trained model to evaluate the efficacy of our model. As an illustration, to make predictions about the user engagement level for the product advertisements on February 23rd, we will train a model build on logs from February 1st until February 22nd. Similarly, to make predictions on February 24th, we will train a model which is build on traffic logs from February 2nd until February 23th. 

\section{Proposed Models}
In this section, we would be discussing the techniques used to model the engagement of a user with product related advertisement, while being responsive to changes in user buying or product selling behavior. 

\subsection{Historic Conversion Rate Feature Model (HCRFM)} 
The first proposed model is an extension of our baseline model in the sense that the model is obtained by adding features which are indicative of the variation in product demand in conjunction with the existing features. The rationale being that the addition of novel features might help the models to better accommodate the effects of changes in user buying or product selling behavior. These features are derived from past aggregate conversion rates at an advertiser level. A higher value of this feature indicates that an advertiser's products are in demand, whereas a low value of this feature indicates the demand for an advertiser's products is on decline.

This new conversion rate feature $\text{cr}_i$ is derived from the function $\text{CR}(a,d)$ that computes the conversion rate for advertiser $'a'$ on the calendar day $'d'$ for the $'ith'$ data point  

\begin{eqnarray}
\text{cr}_i = \text{CR}(a_i,d_i-1) 
\end{eqnarray}
where in,

\begin{itemize}
  \item  $a_i$ denotes the advertiser corresponding to the $i^{th}$ event
  \item $d_i$ denotes the calendar day of the $i^{th}$ event
\end{itemize}

Once such features have been constructed, then the modified $L_2$ regularized logistic loss is used to obtain the coefficient parameter $\w$. 

\begin{eqnarray}
\arg\min_\w \sum_{i=1}^N \log(1+\exp(-y_i\w\cdot (\x_i 
+\log \text{cr}_i  )) + \frac{\lambda}{2}\|\w\|^2
\end{eqnarray} 

\subsection{Time Decay Weighting Model (TDWM)}
To incorporate the effects of changing catalog and user behavior over time, we hypothesize that models built on data from recent past might be a better fit as compared to models built on data from distant past. In our second proposed model (TDWM) the weight of each data point is given by a time decay function. Data points from the recent past are weighted more as compared to data points from the distant past. To compute the model parameters, the loss function of the baseline model (i.e. logistic regression) is slightly altered and is denoted by weighted-negative-log-likelihood  (WNLL). 

We define:  \begin{eqnarray}
\text{WNLL} = \sum_{i=1}^N d(t_i) \log(1+\exp(-y_i\w\cdot\x_i)) + \frac{\lambda}{2}\|\w\|^2,
\end{eqnarray} 

Each data point is weighted by $d(t_i)$, an exponential time decay function with a half life of 5 days. Half-life is defined as the time taken for a data point to reduce its weight by 50\%. The exponential decay function is expressed as, 
 \begin{eqnarray}
d(t_i) = 2^{-\frac{\text{age}(t_i)}{5}}
\end{eqnarray} 
where $t_i$ denotes the time and $\text{age}(t_i)$ is the  difference expressed in days between $t_i$ and a reference time $t_0$.  We chose the half life to be 5 days based on experimental evaluation.

\begin{figure}[h]
	\includegraphics[width=\columnwidth]{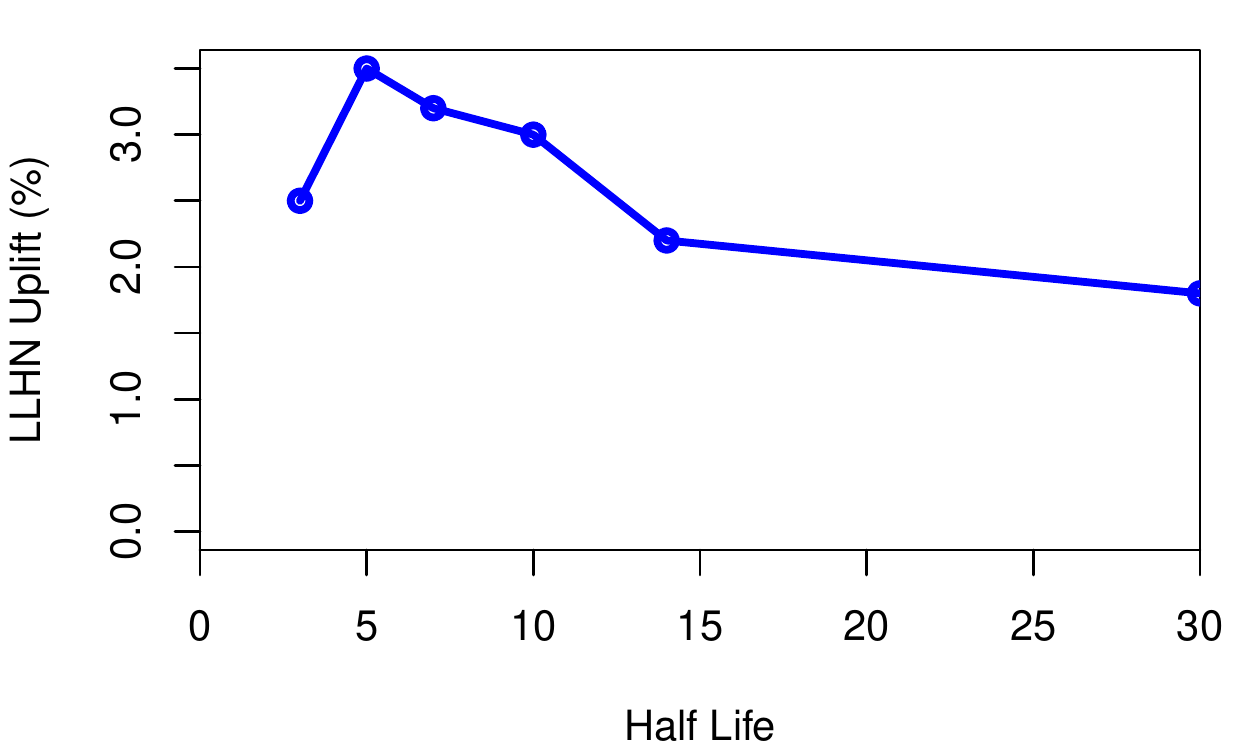}
    \centering
    \caption{Uplift in LLHN as a function of the half life decay.}
    \label{fig:half_life}
\end{figure}

  Figure~\ref{fig:half_life} shows the uplift in LLHN as function of the half life decay. The LLHN uplift is the average of the uplift obtained on five different 7-day periods across all advertisers. The x-axis indicates different half-live values ranging from 3 to 30 days. The y-axis represents the average LLHN-Uplift across five different 7-day periods. As observed, the half-live value of 5 days has the maximum LLHN-Uplift.

\subsubsection{Parameter Estimation} 
We analyze the learning setup proposed in \cite{chapelle2015simple} where limited memory BFGS \cite{liu1989limited,nocedal1980updating} (L-BFGS) is warm-started with stochastic gradient descent \cite{bottou2010large} (SGD). For both algorithms, we multiply the gradient of the loss of each example by $d(t_i)$, where $d(t_i)$ is the weight  associated with the example $i$ computed by the decay function. 

\paragraph{Impact on the regularization parameter}
\label{sec:lambda}

In the case of switching from log loss (NLL, equation-2) to the weighed log loss (WNLL, equation-5) the value of the $\lambda$ hyper-parameter for NLL needs to be adapted to WNLL. To do that, we use the following simple rule that adapts $\lambda$ depending on the value of the importance weights used, i.e. of the average value of the decay function:
\begin{eqnarray}
\lambda_{WNLL} = \lambda_{NLL} \times \frac{\sum_i d(t_i)}{N}
\end{eqnarray}

\subsection{Mixture of Long-Term and Short-Term Model (MLTSTM)}
Our third proposed model (MLTSTM) is a mixture of long-term and Short-Term models. In this model, the prediction is a weighted average of the prediction from two models, a \emph{Short-Term} and a \emph{Long-Term} model. The difference (Short-Term vs Long-Term) lies in the timespan of the data used to train such models. The Short-Term model is trained on data from the recent past whereas the Long-Term model is trained on data from recent as well as distant past. 

The model prediction (denoted by $E(y_i / x_i)$)  is a weighted average of the prediction from two models, a \emph{Short-Term} and a \emph{Long-Term} model.
\begin{eqnarray}
E(y_i / x_i) = \alpha  \times E(y_i / x_i , w_1)  + (1 - \alpha) \times E(y_i / x_i , w_2)
\end{eqnarray} 
where in,

\begin{itemize}
  \item $w_1$ denotes the weight parameter obtained from the Short-Term model
  \item $w_2$ denotes the weight parameter obtained from the Long-Term model  
  \item $E(y_i / x_i , w_1)$ is obtained from the Short-Term model trained on data from recent past 
  \item $E(y_i / x_i , w_2)$ is obtained from the Long-Term model trained on data from recent as well as distant past 
  \item $\alpha$ denotes the average weighting factor and ranges from 0 to 1.
\end{itemize}

\begin{figure}[h]
	\includegraphics[width=\columnwidth]{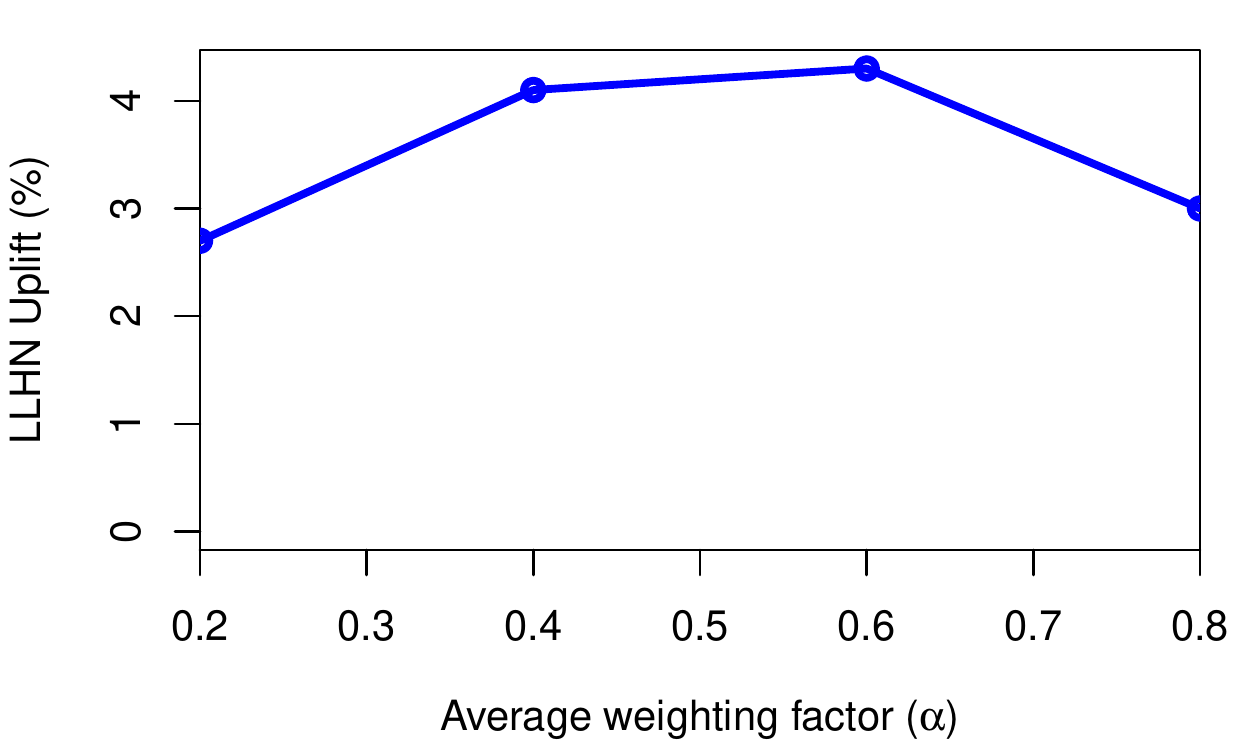}
    \centering
    \caption{LLHN-Uplift as a function of the average weighting factor $\alpha$.}
    \label{fig:shortterm}
\end{figure}
Figure~\ref{fig:shortterm}, shows LLHN-Uplift as a function of the average weighting factor $\alpha$.  The x-axis indicates different values of $\alpha$. The y-axis is the average LLHN-Uplift across five different 7-day periods. As observed, corresponding to $\alpha = 0.6$, we observe the best LLHN-Uplift. 

\section{Experiments and Results}
In this section we will first compare the performance of the proposed models when there is considerable variation in product demand against the performance when there is no considerable variations. We carried out this experiment on a controlled group of advertisers who have experienced variations in product demand. Next, we will evaluate the performance of the proposed models during a period known to have high incidence of variations in product demand, namely \emph{Black Friday}. 
Both experiments were performed offline based on event logs collected by Criteo's Predictive Search (see Section \ref{sec:CPS}). 

\subsection{Evaluating Responsiveness to Variations in Product Demand}
\label{sec:seasonality_comparison}
The purpose of this experiment is to compare the responsiveness of the proposed models when there is considerable variation in product demand against the performance when there is no considerable variations.
 Our hypothesis is that the more pronounced the variation in product demand, the greater the LLHN-Uplift we will observe from the proposed models. 
Because performances vary significantly across different advertisers, we will compare performances only within single advertisers to avoid introducing additional noise.

\begin{table}[h]
  \caption{Traffic volume for the advertisers included in the comparison study of Section~\ref{sec:seasonality_comparison} (daily averages).}
  \label{tab:traffic_volume}
  \centering
    \begin{tabular}{rrrr}
    \toprule
    Advertiser & Events & Sales  & CR (\%) \\
    \midrule
    Advertiser 1	& 21500	& 250	& 1.2 \\
    Advertiser 2	& 5500	& 100	& 1.8 \\
    Advertiser 3	& 850	& 20		& 2.4 \\
    Advertiser 4	& 14550	& 1150	& 7.9 \\
    Advertiser 5	& 5500	& 100	& 1.8 \\
    \bottomrule
    \end{tabular}
\end{table}

\begin{table}[t]
  \caption{
  LLHN-Uplift of the proposed models for advertiser's different levels of variation in product demand}
  \label{tab:llhcvn_model_seasonality}
  \centering
  \begin{tabular}{lrrrrr}
    Model				&  Adv. 1		&  Adv. 2	&  Adv. 3	& Adv. 4	&  Adv. 5  \\
    \midrule
        \noalign{\smallskip}
    	& \multicolumn{5}{c}{Extreme Variation} \\
         \cmidrule(lr){2-6}
   HCRFM 	&  54.3			&  81.6 		&  180.1 		&  13.8 		&  229.4	\\
    TDWM 	&  70.5 			&  47.3 		&  257.5 		&  20.1 		&  361.6  \\
    MLTSTM	& \textbf{90.1} 		&\textbf{124.6} & \textbf{340.2}	&  \textbf{25.6}	&  \textbf{596.6}  \\
        \noalign{\smallskip}
     	& \multicolumn{5}{c}{Average Variation} \\
         \cmidrule(lr){2-6}
    HCRFM 	&  2.4			&  \textbf{0.9}	&  -6.0		&  56.2		&  37.2  \\
    TDWM	&  7.0			&  -2.7		&  \textbf{22.0}	&  198.0		&  73.4 \\
    MLTSTM	& \textbf{7.7}		&  -0.7		&  11.1		& \textbf{217.3}	&  \textbf{79.1}  \\
        \noalign{\smallskip}
     	& \multicolumn{5}{c}{Moderate Variation} \\
         \cmidrule(lr){2-6}
   HCRFM 	&  0.1			&  \textbf{-1.7}	&  \textbf{1.9}	&  -18.2		&  \textbf{-1.2}  \\
    TDWM	&  \textbf{6.2}		&  -9.8		&  -9.6		&  \textbf{6.7}	&  -6.9  \\
    MLTSTM	& 4.2				&  -8.1		&  -14.4		&  -4.8		&  -5.6  \\
         \bottomrule
  \end{tabular}
\end{table}

\begin{figure*}[t]
    \centering
    \begin{subfigure}{0.49\textwidth}
        \includegraphics[width=\textwidth]{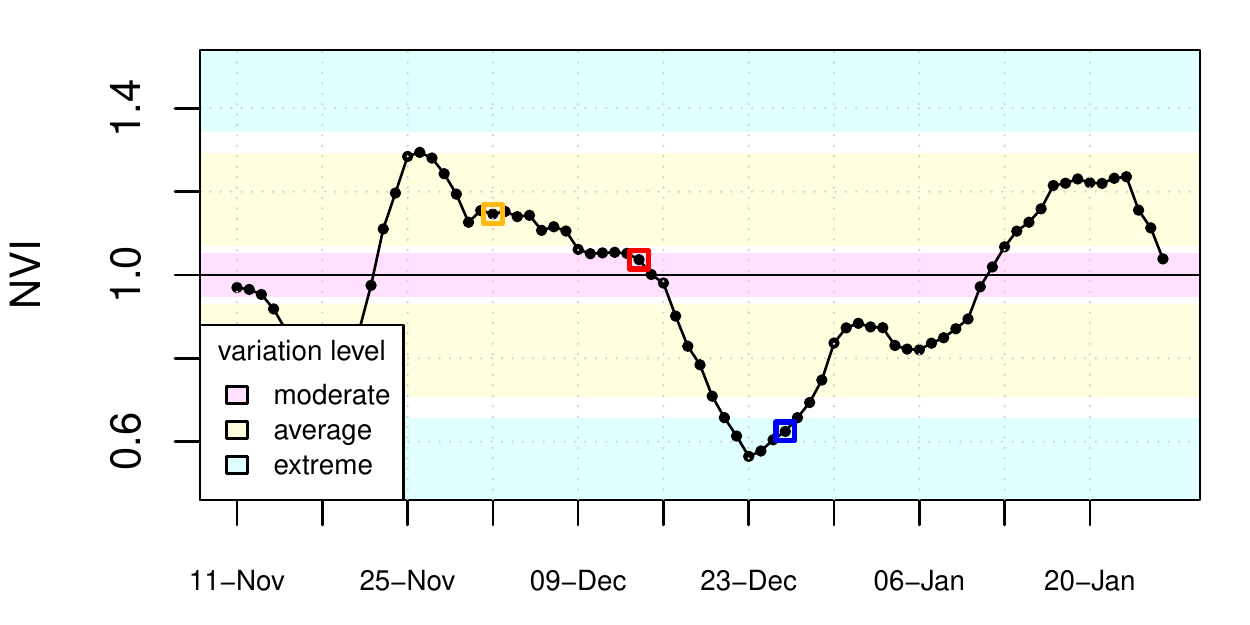}
        \caption{Advertiser 1.}
        \label{fig:advertiser1}
    \end{subfigure}
    \begin{subfigure}{0.49\textwidth}
        \includegraphics[width=\textwidth]{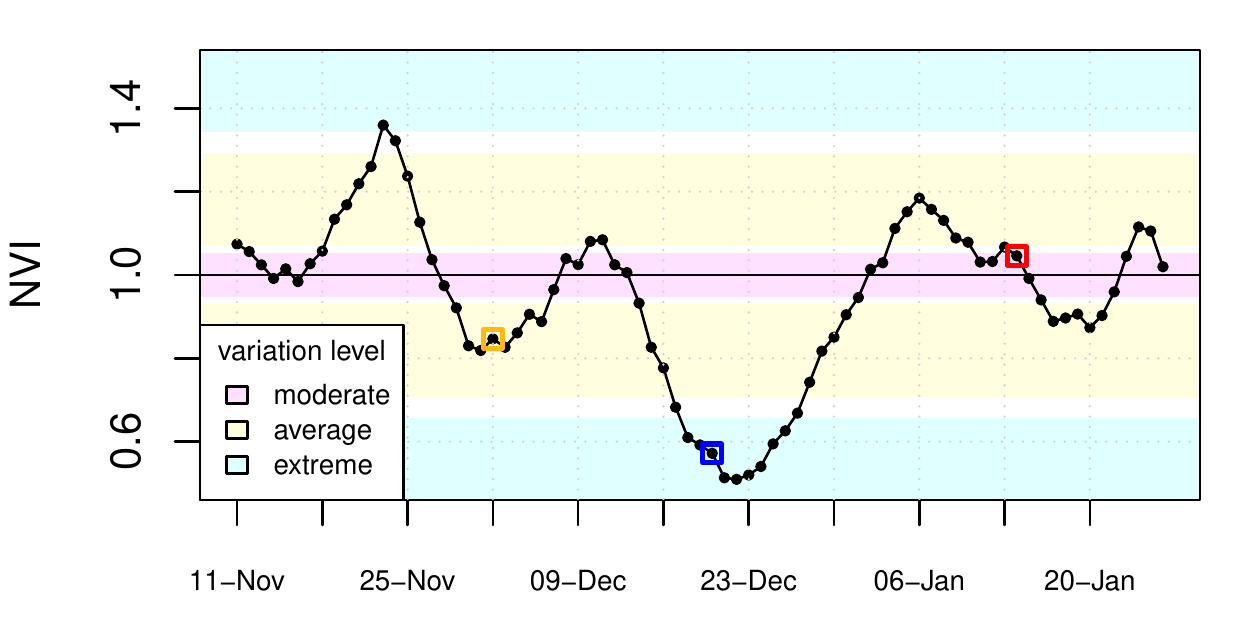}
        \caption{Advertiser 2.}
        \label{fig:advertiser2}
    \end{subfigure}
    \begin{subfigure}{0.49\textwidth}
        \includegraphics[width=\textwidth]{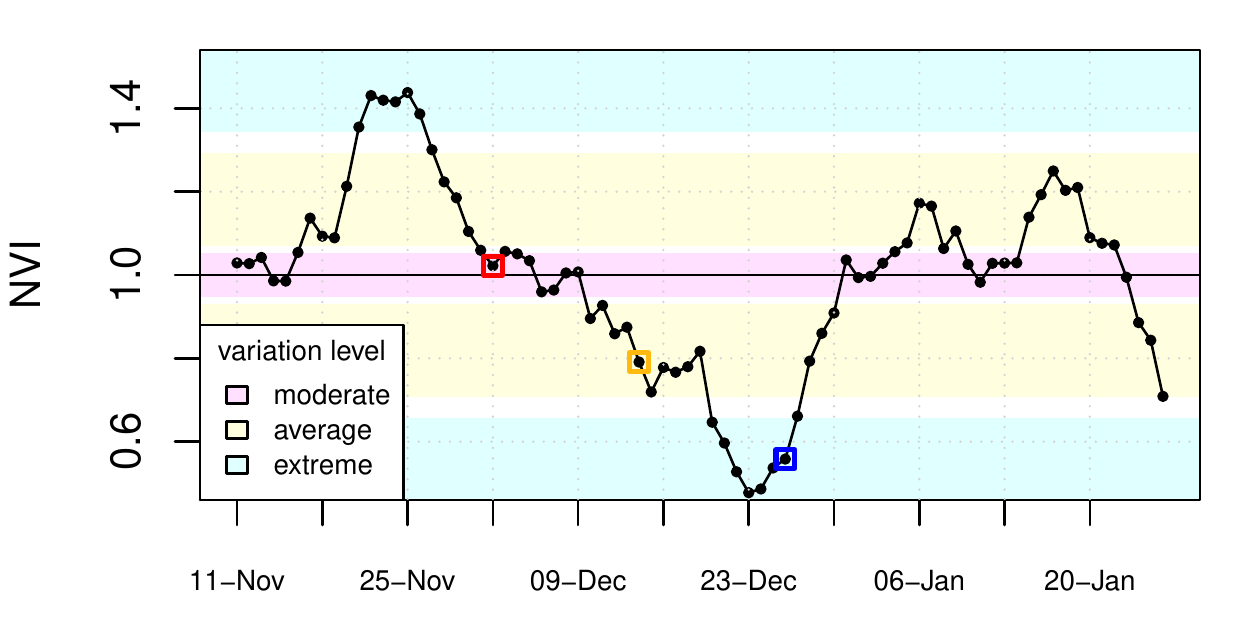}
        \caption{Advertiser 3.}
        \label{fig:advertiser3}
    \end{subfigure}
    \begin{subfigure}{0.49\textwidth}
        \includegraphics[width=\textwidth]{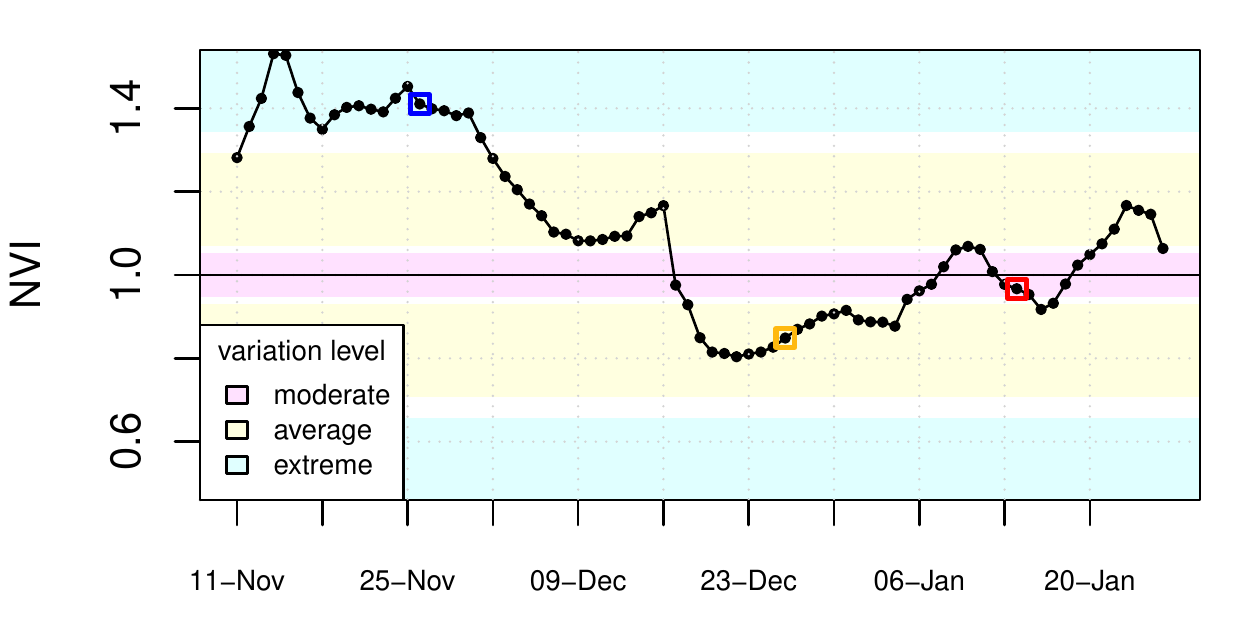}
        \caption{Advertiser 4.}
        \label{fig:advertiser4}
    \end{subfigure}
    \begin{subfigure}{0.49\textwidth}
        \includegraphics[width=\textwidth]{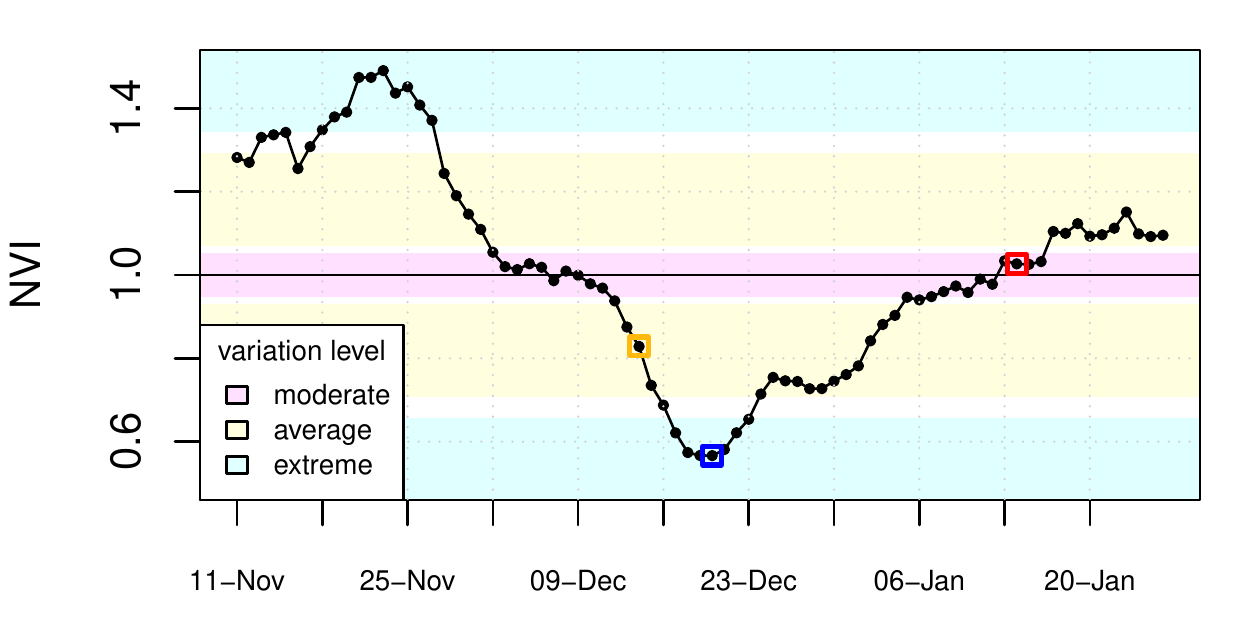}
        \caption{Advertiser 5.}
        \label{fig:advertiser5}
    \end{subfigure}
    \caption{Advertisers normalized variation index (NVI) over time.}
    \label{fig:partner_seasonality_levels}
\end{figure*}

The aim of this experiment is to compare the performance of the proposed models across periods with different levels of variations in product demand.
To carry out this experiment we need a working definition of level of variation in product demand.

We define a metric to measure the level of variation in product demand for an advertiser on a given (narrow) period of time as the ratio between the \emph{conversion rate} of two periods, one shorter period in the numerator and a longer period in the denominator. We call this metric the \textbf{Normalized Variation Index (NVI)}.
For the study reported, in this manuscript our NVI metric relied on a seven-day period in the numerator and its preceding 30-day period in  the denominator.
More formally:
\begin{eqnarray}
    V^a_d  = \frac{\text{CR}^a_{[d,d+6]}}{\text{CR}^a_ {[d-30,d-1]}}
    \label{eq:seasonality_metric}
\end{eqnarray}
where $V^a_d$ denotes the NVI for advertiser $a$ on a period beginning on day $d$, 
and $\text{CR}^a_{[d_i,d_j]}$ is the \emph{conversion rate} for advertiser $a$ over the period $[d_i,d_j]$.

Intuitively, the NVI metric indicates the degree at which an advertiser's conversion rate over a narrow period of time has deviated from a historic conversion rate, which is represented by the normalizing factor in the denominator of the NVI definition. 
According to this definition, an NVI value close to '1' indicates a period with little or no variation, an NVI value grater than '1' indicates a surge in product demand and an NVI value lower than '1' indicates a drop in product demand. Furthermore, the farther away an NVI value is from '1', it indicates the more extreme the variation in product demand is.

We rely on the NVI metric to select a set of test cases appropriate for this study from the CPS event logs. 
We wanted to test our hypothesis that the proposed models will be more accurate than the baseline model during periods of extreme variation in product demand. However, we also wanted to test how the proposed models compare to the baseline on periods of average and moderate variation. In order to carry out these tests we defined three different conditions of interests for this study based on how extreme an NVI value is. 
The three conditions are \emph{extreme} variation, \emph{average} variation and \emph{moderate} variation.
These conditions were defined as follows:
\begin{eqnarray}
	f(a,d) = \left\{
	  \begin{array}{ll}
	    \text{moderate}	& : | 1 - V^a_d | <= 0.05 \\
	    \text{average}	& : 0.07 <= | 1 - V^a_d | <= 0.29 \\
	    \text{extreme}	& : | 1 - V^a_d | >= 0.34 \\
	  \end{array}
	\right.
    \label{eq:seasonality_condition}
\end{eqnarray}
Where $f(a,d)$ denotes the \emph{level of variation in product demand} for advertiser $a$ during period $d$, $V^a_d$ is the \emph{NVI} of advertiser $a$ over period $d$ as defined by equation (\ref{eq:seasonality_metric}). The four range boundary constants 0.05, 0.07,  0.29, and 0.34 were chosen based on a distribution of NVI values for all advertisers and all periods within a 3 months worth of events logs data. These boundaries corresponded to the quantiles 0.20, 0.25, 0.75, and 0.80 respectively.

We then scanned our events logs to select a set of advertisers that have experienced periods in all three different levels of variation in product demand. While selecting advertisers we also procured to obtain a diverse set of advertisers, which are representative of different sale markets and traffic volume levels.  Table~\ref{tab:traffic_volume}, lists the daily average number of events and conversions for each one of the selected advertisers for this study.

Figure~\ref{fig:partner_seasonality_levels}, plots the changes in NVI over time for the five advertisers selected for this study. The plots also identify the three regions corresponding to the three conditions of interest for this study, \emph{extreme} (blue), \emph{average} (yellow) and \emph{moderate} (pink). 
For each advertiser, the three time periods selected to represent the three conditions of interest for this study are indicated by a small square on top of the NVI plot. For example, we can observe that for Advertiser 1 (Figure~\ref{fig:advertiser1}), the variation on product demand is almost neutral (NVI value close to 1) during the first few weeks of December up to December 16. Then the product demand starts to drop until reaching an extreme NVI level of 0.6 on December 23. This means that during the 7-day period starting on December 23, CR is down at a 60\% of this advertiser's historic CR. 

Table~\ref{tab:llhcvn_model_seasonality}, compares the proposed models performance for five advertisers on periods of extreme, average, and moderate variation in product demand. 
The table indicates the LLHN-Uplift of the proposed models compared to the baseline.
We observe that for most advertisers, the more extreme the variation in product demand the better the performance of the proposed models in relation to the baseline. 
We also observe that most of the time the performance of the MLTSTM model dominates under the extreme and average conditions and HCRFM model does best under the moderate conditions.
Sometimes, under moderate conditions the performance of the proposed models is worse than the baseline. 
However, the loss in performance during moderate conditions are less pronounced than the gains in performance during extreme conditions. 
Both conditions, moderate and extreme, are equally rare under this experimental settings.

\subsection{Black Friday season}
The constraints imposed  by the study described in Section~ \ref{sec:seasonality_comparison}, determined that only a small sample of advertisers were eligible, which might bring into question the significance of its results. 
Experiments discussed in this section, aim to compensate for the limitations of the above study by bringing a more comprehensive sample of advertisers at the expense of performing a direct comparison across different levels of variation in product demand.

In this study we evaluated the proposed models on all of U.S. advertisers during a 7-day period that included Black Friday.  The rationale for this study is the belief that a significant fraction of these advertisers for the U.S. market will experience extreme variations during the selected period. 

Table~\ref{tab:black_friday_aggregated}, shows the LLHN-Uplift of the proposed models when aggregating across all U.S. advertisers during the selected period. Models TDWM and MLTSTM show a positive uplift, although a modest one. The largest uplift was 3.20\% for model MLTSTM. Although this uplift might seem small, we should take into account that the impact of the proposed models might have been diluted when considering it in the aggregated context. The proposed models were conceived specifically to improve the prediction performance during periods of steep variations in product demand. However, our aggregated sample includes several advertisers that have not experienced variations in product demand.

\begin{table}[h]
  \caption{
  LLHN-Uplift of the proposed models during a \emph{Black Friday} week over all US advertisers..
  }
  \label{tab:black_friday_aggregated}
  \centering
    \begin{tabular}{rr}
    \toprule
    Model & LLHN-Uplift (\%) \\
    \midrule
    HCRFM & 0.2 \\
    TDWM & 2.9 \\
   MLTSTM & \textbf{3.2} \\
    \bottomrule
    \end{tabular}
\end{table}

\begin{table}[h]
  \caption{
   Performance (LLHN-Uplift) of the proposed models during a \emph{Black Friday} week for the top US advertisers.}
  \label{tab:black_friday_top_advertisers}
  \centering
    \begin{tabular}{ccrrr}
    \toprule
    Advertiser & Extremeness & HCRFM & TDWM & MLTSTM \\
    \midrule
    1 & 1.81	& 62.1		& 239.4	& \textbf{409.2} \\
    2 & 0.94	& -2.9 		& 117.6	& \textbf{136.2} \\
    3 & 0.41	& 0.7			& 15.7	& \textbf{18.2} \\
    4 & 0.40	& \textbf{0.5} 	& -2.7	& -7.2 \\
    5 & 0.36	& 0.3			& \textbf{5.9}	& 1.2 \\
    6 & 0.25	& 0.0			& \textbf{7.4} & 6.7 \\
    7 & 0.17	& -0.6		& 10.3	& \textbf{10.8} \\
    8 & 0.12 	& \textbf{0.9}	& -0.1 	& -1.0 \\
    9 & 0.11	& \textbf{0.3} 	& -0.8 	& -0.6 \\
    10 & 0.02 	& 0.5 		& 2.3 	& \textbf{4.3} \\
    \bottomrule
    \end{tabular}
\end{table}

Table~\ref{tab:black_friday_top_advertisers}, on the other hand, presents a more focused analysis. It shows the performance of the proposed models for each of our top ten U.S. advertisers in terms of traffic volume. 
The table lists advertisers in \emph{NVI extremeness} decreasing order. 
Here, NVI extremeness is  $| 1 - V^a_d |$ where $V^a_d$ is the NVI metric as defined in equation  (\ref{eq:seasonality_metric}).
In general, we can observe that the more extreme the NVI the better the performance of the proposed models are relative to the performance of the baseline model.

\begin{figure*}[h]
    \centering
    \begin{subfigure}{0.32\textwidth}
        \includegraphics[width=\textwidth]{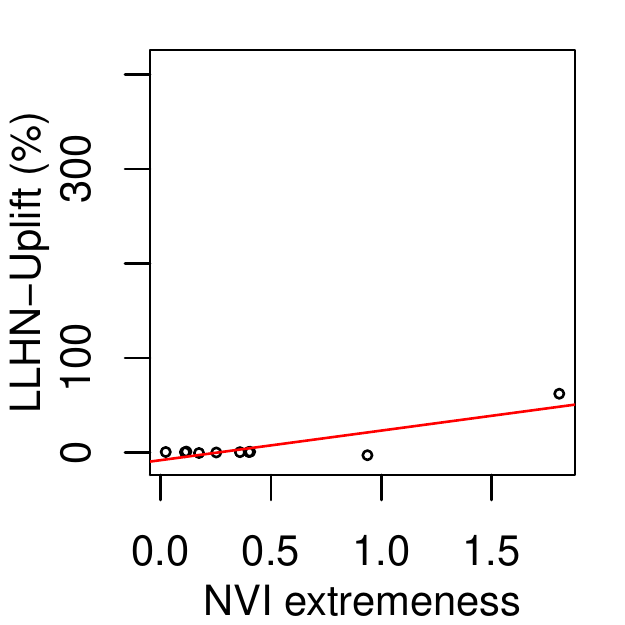}
        \caption{HCRFM.}
    \end{subfigure}
    \begin{subfigure}{0.32\textwidth}
        \includegraphics[width=\textwidth]{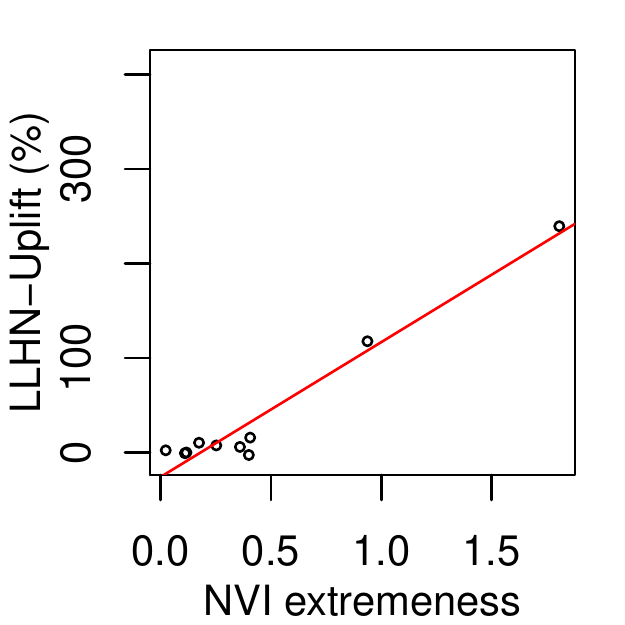}
        \caption{TDWM.}
    \end{subfigure}
    \begin{subfigure}{0.32\textwidth}
        \includegraphics[width=\textwidth]{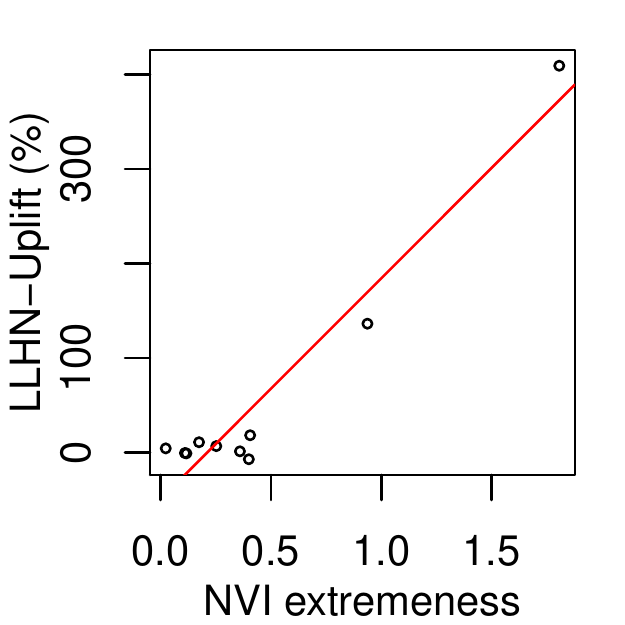}
        \caption{MLTSTM.}
    \end{subfigure}
    \caption{Relationship between normalized variation index (NVI) and LLHN-Uplift during Black Friday period for the 10 advertisers.}
    \label{fig:BF_upliftXextremeness}
\end{figure*}

Alternatively, in Figure \ref{fig:BF_upliftXextremeness}, we present individual scatter plots for the proposed models. For each proposed model, we present the NVI extremeness on the x-axis and LHHN-Uplift on the Y-axis. As depicted in the graphs, we observe that for model MLTSTM, LLHN-Uplift is very high for the advertiser which has high NVI extremeness values. Similar, is the trend observed for the scatter plot obtained from the model TDWM. Such high LLHN uplifts can be attributed to the fact that these models i.e. TDWM and MLTSTM are better able to handle variations in user buying or product selling behavior as compared to the HCRFM model.

\section{Conclusion}
In this work, we proposed three CR prediction models which are robust to variations in product demand. In the first model (HCRFM) we extended the baseline model by adding novel features which are indicative of variation in product demand. The second technique (TDWM) consists of a model in conjunction with importance weighting \cite{vasile2016cost}, where in data from the recent past is weighted more as compared to data from the distant past during model training, and lastly we propose mixture models (MLTSTM) i.e. models consisting of our original model along with a model trained on more recent data.

In this work we also defined a novel metric to measure the variation in the product demand of an advertiser on a given (narrow) period of time and observe the different variations as exhibited by multiple advertisers over time. Using this  metric, we observe the performance of the proposed models over an advertiser\'s different variation conditions i.e. moderate, extreme and average. We observed that for most advertisers, the more extreme the product demand variation condition the better the performance of the proposed models w.r.t the baseline model.

Further, to evaluate the performance of the proposed models during periods of high variation, we evaluated the performance of the proposed models on all of U.S.  advertisers during a 7-day period that included Black Friday (high product variation season) and observe the positive uplift in LLHN obtained by using the TDWM and MLTSTM models. We also analyze the performance of the proposed models on each of our top ten U.S. advertisers during the Black-Friday period and observed that maximum uplift was associated with the advertiser which experiences the highest product demand variation. 

Our results demonstrate that our proposed models (i.e. HCRFM, TDWM and MLTSTM) can help us to better model the engagement between the user and product related advertisement. These models achieve it by weighing data samples in the recent past a bit higher as compared to the data samples from the distant past. Such modeling can then help us to avoid over-predicting the engagement levels when there is a drop in product demand  and to avoid under-predicting the engagement levels when there is a surge in product demand. In future, we would like to explore non-linear modeling techniques such as Gradient Boosting Decision Trees (GBDTs) and Convolution Neural Networks (CNNs) to compute the engagement between the user and product advertisement subject to the constraints (i.e. fast update of the model parameters given new data, computing inferences for billion of products in minute fraction of time, model training on extremely sparse datasets).

\bibliographystyle{ACM-Reference-Format}
\bibliography{paper}


\begin{thebibliography}{20}


\ifx \showCODEN    \undefined \def \showCODEN     #1{\unskip}     \fi
\ifx \showDOI      \undefined \def \showDOI       #1{#1}\fi
\ifx \showISBNx    \undefined \def \showISBNx     #1{\unskip}     \fi
\ifx \showISBNxiii \undefined \def \showISBNxiii  #1{\unskip}     \fi
\ifx \showISSN     \undefined \def \showISSN      #1{\unskip}     \fi
\ifx \showLCCN     \undefined \def \showLCCN      #1{\unskip}     \fi
\ifx \shownote     \undefined \def \shownote      #1{#1}          \fi
\ifx \showarticletitle \undefined \def \showarticletitle #1{#1}   \fi
\ifx \showURL      \undefined \def \showURL       {\relax}        \fi
\providecommand\bibfield[2]{#2}
\providecommand\bibinfo[2]{#2}
\providecommand\natexlab[1]{#1}
\providecommand\showeprint[2][]{arXiv:#2}

\bibitem[\protect\citeauthoryear{Agarwal, Agrawal, Khanna, and Kota}{Agarwal
  et~al\mbox{.}}{2010}]%
        {agarwal2010estimating}
\bibfield{author}{\bibinfo{person}{Deepak Agarwal}, \bibinfo{person}{Rahul
  Agrawal}, \bibinfo{person}{Rajiv Khanna}, {and} \bibinfo{person}{Nagaraj
  Kota}.} \bibinfo{year}{2010}\natexlab{}.
\newblock \showarticletitle{Estimating rates of rare events with multiple
  hierarchies through scalable log-linear models}. In
  \bibinfo{booktitle}{\emph{Proceedings of the 16th ACM SIGKDD international
  conference on Knowledge discovery and data mining}}. ACM,
  \bibinfo{pages}{213--222}.
\newblock


\bibitem[\protect\citeauthoryear{Bottou}{Bottou}{2010}]%
        {bottou2010large}
\bibfield{author}{\bibinfo{person}{L{\'e}on Bottou}.}
  \bibinfo{year}{2010}\natexlab{}.
\newblock \showarticletitle{Large-scale machine learning with stochastic
  gradient descent}.
\newblock In \bibinfo{booktitle}{\emph{Proceedings of COMPSTAT'2010}}.
  \bibinfo{publisher}{Springer}, \bibinfo{pages}{177--186}.
\newblock


\bibitem[\protect\citeauthoryear{Chapelle, Manavoglu, and Rosales}{Chapelle
  et~al\mbox{.}}{2015}]%
        {chapelle2015simple}
\bibfield{author}{\bibinfo{person}{Olivier Chapelle}, \bibinfo{person}{Eren
  Manavoglu}, {and} \bibinfo{person}{Romer Rosales}.}
  \bibinfo{year}{2015}\natexlab{}.
\newblock \showarticletitle{Simple and scalable response prediction for display
  advertising}.
\newblock \bibinfo{journal}{\emph{ACM Transactions on Intelligent Systems and
  Technology (TIST)}} \bibinfo{volume}{5}, \bibinfo{number}{4}
  (\bibinfo{year}{2015}), \bibinfo{pages}{61}.
\newblock


\bibitem[\protect\citeauthoryear{Chen and Yan}{Chen and Yan}{2012}]%
        {chen2012position}
\bibfield{author}{\bibinfo{person}{Ye Chen} {and} \bibinfo{person}{Tak~W Yan}.}
  \bibinfo{year}{2012}\natexlab{}.
\newblock \showarticletitle{Position-normalized click prediction in search
  advertising}. In \bibinfo{booktitle}{\emph{Proceedings of the 18th ACM SIGKDD
  international conference on Knowledge discovery and data mining}}. ACM,
  \bibinfo{pages}{795--803}.
\newblock


\bibitem[\protect\citeauthoryear{Cheng and Cant{\'u}-Paz}{Cheng and
  Cant{\'u}-Paz}{2010}]%
        {cheng2010personalized}
\bibfield{author}{\bibinfo{person}{Haibin Cheng} {and} \bibinfo{person}{Erick
  Cant{\'u}-Paz}.} \bibinfo{year}{2010}\natexlab{}.
\newblock \showarticletitle{Personalized click prediction in sponsored search}.
  In \bibinfo{booktitle}{\emph{Proceedings of the third ACM international
  conference on Web search and data mining}}. ACM, \bibinfo{pages}{351--360}.
\newblock


\bibitem[\protect\citeauthoryear{Fain and Pedersen}{Fain and Pedersen}{2006}]%
        {fain2006sponsored}
\bibfield{author}{\bibinfo{person}{Daniel~C Fain} {and} \bibinfo{person}{Jan~O
  Pedersen}.} \bibinfo{year}{2006}\natexlab{}.
\newblock \showarticletitle{Sponsored search: A brief history}.
\newblock \bibinfo{journal}{\emph{Bulletin of the Association for Information
  Science and Technology}} \bibinfo{volume}{32}, \bibinfo{number}{2}
  (\bibinfo{year}{2006}), \bibinfo{pages}{12--13}.
\newblock


\bibitem[\protect\citeauthoryear{He, Pan, Jin, Xu, Liu, Xu, Shi, Atallah,
  Herbrich, Bowers, et~al\mbox{.}}{He et~al\mbox{.}}{2014}]%
        {he2014practical}
\bibfield{author}{\bibinfo{person}{Xinran He}, \bibinfo{person}{Junfeng Pan},
  \bibinfo{person}{Ou Jin}, \bibinfo{person}{Tianbing Xu}, \bibinfo{person}{Bo
  Liu}, \bibinfo{person}{Tao Xu}, \bibinfo{person}{Yanxin Shi},
  \bibinfo{person}{Antoine Atallah}, \bibinfo{person}{Ralf Herbrich},
  \bibinfo{person}{Stuart Bowers}, {et~al\mbox{.}}}
  \bibinfo{year}{2014}\natexlab{}.
\newblock \showarticletitle{Practical lessons from predicting clicks on ads at
  facebook}. In \bibinfo{booktitle}{\emph{Proceedings of the Eighth
  International Workshop on Data Mining for Online Advertising}}. ACM,
  \bibinfo{pages}{1--9}.
\newblock


\bibitem[\protect\citeauthoryear{Jiang}{Jiang}{2016}]%
        {jiang2016research}
\bibfield{author}{\bibinfo{person}{Zilong Jiang}.}
  \bibinfo{year}{2016}\natexlab{}.
\newblock \showarticletitle{Research on ctr prediction for contextual
  advertising based on deep architecture model}.
\newblock \bibinfo{journal}{\emph{Journal of Control Engineering and Applied
  Informatics}} \bibinfo{volume}{18}, \bibinfo{number}{1}
  (\bibinfo{year}{2016}), \bibinfo{pages}{11--19}.
\newblock


\bibitem[\protect\citeauthoryear{Juan, Zhuang, Chin, and Lin}{Juan
  et~al\mbox{.}}{2016}]%
        {juan2016field}
\bibfield{author}{\bibinfo{person}{Yuchin Juan}, \bibinfo{person}{Yong Zhuang},
  \bibinfo{person}{Wei-Sheng Chin}, {and} \bibinfo{person}{Chih-Jen Lin}.}
  \bibinfo{year}{2016}\natexlab{}.
\newblock \showarticletitle{Field-aware factorization machines for CTR
  prediction}. In \bibinfo{booktitle}{\emph{Proceedings of the 10th ACM
  Conference on Recommender Systems}}. ACM, \bibinfo{pages}{43--50}.
\newblock


\bibitem[\protect\citeauthoryear{King, Atkins, and Schwarz}{King
  et~al\mbox{.}}{2007}]%
        {king2007internet}
\bibfield{author}{\bibinfo{person}{Mervyn King}, \bibinfo{person}{Jill Atkins},
  {and} \bibinfo{person}{Michael Schwarz}.} \bibinfo{year}{2007}\natexlab{}.
\newblock \showarticletitle{Internet advertising and the generalized
  second-price auction: Selling billions of dollars worth of keywords}.
\newblock \bibinfo{journal}{\emph{The American economic review}}
  \bibinfo{volume}{97}, \bibinfo{number}{1} (\bibinfo{year}{2007}),
  \bibinfo{pages}{242--259}.
\newblock


\bibitem[\protect\citeauthoryear{Liu and Nocedal}{Liu and Nocedal}{1989}]%
        {liu1989limited}
\bibfield{author}{\bibinfo{person}{Dong~C Liu} {and} \bibinfo{person}{Jorge
  Nocedal}.} \bibinfo{year}{1989}\natexlab{}.
\newblock \showarticletitle{On the limited memory BFGS method for large scale
  optimization}.
\newblock \bibinfo{journal}{\emph{Mathematical programming}}
  \bibinfo{volume}{45}, \bibinfo{number}{1} (\bibinfo{year}{1989}),
  \bibinfo{pages}{503--528}.
\newblock


\bibitem[\protect\citeauthoryear{McCullagh}{McCullagh}{1984}]%
        {mccullagh1984generalized}
\bibfield{author}{\bibinfo{person}{Peter McCullagh}.}
  \bibinfo{year}{1984}\natexlab{}.
\newblock \showarticletitle{Generalized linear models}.
\newblock \bibinfo{journal}{\emph{European Journal of Operational Research}}
  \bibinfo{volume}{16}, \bibinfo{number}{3} (\bibinfo{year}{1984}),
  \bibinfo{pages}{285--292}.
\newblock


\bibitem[\protect\citeauthoryear{McMahan, Holt, Sculley, Young, Ebner, Grady,
  Nie, Phillips, Davydov, Golovin, et~al\mbox{.}}{McMahan
  et~al\mbox{.}}{2013}]%
        {mcmahan2013ad}
\bibfield{author}{\bibinfo{person}{H~Brendan McMahan}, \bibinfo{person}{Gary
  Holt}, \bibinfo{person}{David Sculley}, \bibinfo{person}{Michael Young},
  \bibinfo{person}{Dietmar Ebner}, \bibinfo{person}{Julian Grady},
  \bibinfo{person}{Lan Nie}, \bibinfo{person}{Todd Phillips},
  \bibinfo{person}{Eugene Davydov}, \bibinfo{person}{Daniel Golovin},
  {et~al\mbox{.}}} \bibinfo{year}{2013}\natexlab{}.
\newblock \showarticletitle{Ad click prediction: a view from the trenches}. In
  \bibinfo{booktitle}{\emph{Proceedings of the 19th ACM SIGKDD international
  conference on Knowledge discovery and data mining}}. ACM,
  \bibinfo{pages}{1222--1230}.
\newblock


\bibitem[\protect\citeauthoryear{Nocedal}{Nocedal}{1980}]%
        {nocedal1980updating}
\bibfield{author}{\bibinfo{person}{Jorge Nocedal}.}
  \bibinfo{year}{1980}\natexlab{}.
\newblock \showarticletitle{Updating quasi-Newton matrices with limited
  storage}.
\newblock \bibinfo{journal}{\emph{Mathematics of computation}}
  \bibinfo{volume}{35}, \bibinfo{number}{151} (\bibinfo{year}{1980}),
  \bibinfo{pages}{773--782}.
\newblock


\bibitem[\protect\citeauthoryear{Regelson and Fain}{Regelson and Fain}{2006}]%
        {regelson2006predicting}
\bibfield{author}{\bibinfo{person}{Moira Regelson} {and} \bibinfo{person}{D
  Fain}.} \bibinfo{year}{2006}\natexlab{}.
\newblock \showarticletitle{Predicting click-through rate using keyword
  clusters}. In \bibinfo{booktitle}{\emph{Proceedings of the Second Workshop on
  Sponsored Search Auctions}}, Vol.~\bibinfo{volume}{9623}.
\newblock


\bibitem[\protect\citeauthoryear{Rendle}{Rendle}{2010}]%
        {rendle2010factorization}
\bibfield{author}{\bibinfo{person}{Steffen Rendle}.}
  \bibinfo{year}{2010}\natexlab{}.
\newblock \showarticletitle{Factorization machines}. In
  \bibinfo{booktitle}{\emph{Data Mining (ICDM), 2010 IEEE 10th International
  Conference on}}. IEEE, \bibinfo{pages}{995--1000}.
\newblock


\bibitem[\protect\citeauthoryear{Richardson, Dominowska, and Ragno}{Richardson
  et~al\mbox{.}}{2007}]%
        {richardson2007predicting}
\bibfield{author}{\bibinfo{person}{Matthew Richardson}, \bibinfo{person}{Ewa
  Dominowska}, {and} \bibinfo{person}{Robert Ragno}.}
  \bibinfo{year}{2007}\natexlab{}.
\newblock \showarticletitle{Predicting clicks: estimating the click-through
  rate for new ads}. In \bibinfo{booktitle}{\emph{Proceedings of the 16th
  international conference on World Wide Web}}. ACM, \bibinfo{pages}{521--530}.
\newblock


\bibitem[\protect\citeauthoryear{Vasile, Lefortier, and Chapelle}{Vasile
  et~al\mbox{.}}{2016}]%
        {vasile2016cost}
\bibfield{author}{\bibinfo{person}{Flavian Vasile}, \bibinfo{person}{Damien
  Lefortier}, {and} \bibinfo{person}{Olivier Chapelle}.}
  \bibinfo{year}{2016}\natexlab{}.
\newblock \showarticletitle{Cost-sensitive learning for utility optimization in
  online advertising auctions}.
\newblock \bibinfo{journal}{\emph{arXiv preprint arXiv:1603.03713}}
  (\bibinfo{year}{2016}).
\newblock


\bibitem[\protect\citeauthoryear{Zeff and Aronson}{Zeff and Aronson}{1999}]%
        {zeff1999advertising}
\bibfield{author}{\bibinfo{person}{Robbin~Lee Zeff} {and}
  \bibinfo{person}{Bradley Aronson}.} \bibinfo{year}{1999}\natexlab{}.
\newblock \bibinfo{booktitle}{\emph{Advertising on the Internet}}.
\newblock \bibinfo{publisher}{John Wiley \& Sons, Inc.}
\newblock


\bibitem[\protect\citeauthoryear{Zhang, Dai, Xu, Feng, Wang, Bian, Wang, and
  Liu}{Zhang et~al\mbox{.}}{2014}]%
        {zhang2014sequential}
\bibfield{author}{\bibinfo{person}{Yuyu Zhang}, \bibinfo{person}{Hanjun Dai},
  \bibinfo{person}{Chang Xu}, \bibinfo{person}{Jun Feng},
  \bibinfo{person}{Taifeng Wang}, \bibinfo{person}{Jiang Bian},
  \bibinfo{person}{Bin Wang}, {and} \bibinfo{person}{Tie-Yan Liu}.}
  \bibinfo{year}{2014}\natexlab{}.
\newblock \showarticletitle{Sequential Click Prediction for Sponsored Search
  with Recurrent Neural Networks.}. In \bibinfo{booktitle}{\emph{AAAI}}.
  \bibinfo{pages}{1369--1375}.
\newblock


\end{thebibliography}

\end{document}